\documentstyle[12pt,epsfig]{article}

\input{tcilatex}

\begin{document}

\title{Family Nonuniversal $Z^{\prime}$ and $b\rightarrow s\gamma $ decay}
\author{Ramazan SEVER\thanks{%
Orta Dogu Teknik Universitesi, Ankara- Turkey} \\
Aytekin AYDEMIR\thanks{%
Mersin Universitesi, Icel- Turkey}}
\date{\today }

\thispagestyle{empty} \vskip1.5cm

\baselineskip=15pt
\begin{titlepage}
\thispagestyle{empty}
\maketitle
\begin{abstract}
\thispagestyle{empty}

\baselineskip=15pt

We have calculated the branching ratio and CP asymmetry of
$B\rightarrow X_s \gamma$ decay within the family--nonuniversal
$Z^{\prime}$ models. We have established certain bounds on the
model parameters using the present experimental bounds. We also
comment on the role of family--nonuniversality in the
hadronic decay modes of the $B$ meson.

\end{abstract}


\end{titlepage}
\newpage

\section{Introduction}

The impressive agreement between the standard electroweak theory (SM) and
experiment is not sufficient to make it the fundamental theory of nature up
to very high energies. This follows mainly from the instability of its Higgs
sector under radiative corrections. The supersymmetry (SUSY) is thus one of
the remedies which can cure this hierarchy problem. However, the minimal
supersymmetric model (MSSM) itself has a hierarchy problem concerning the
natural scale of the Higgsino mass parameter ($\mu $ problem). This problem
is solvable in various frameworks each of which introducing a certain SM
singlets to dynamically generate the $\mu $ parameter. The $Z^{\prime }$
models form a viable candidate model to extend the MSSM in order to solve
the $\mu $ problem \cite{Cvetic:1997wu}. The vacua that suppress the $%
Z-Z^{\prime }$ mixing have been found in \cite{cvetic1} with further
improvements concerning the radiative effects \cite{suematsu}. The effects
of such extra $Z$ bosons on precision observables have been analyzed in
detail in \cite{langacker}. Moreover, several collider signatures have been
analyzed for tree level Higgs sector \cite{pak}.

Recently there have been a revived interest in the flavor changing neutral
current phenomena \cite{paul} in such models \cite{plum,hiller}. Given the
recent indications for a light $Z^{\prime}$ boson \cite{lightZp}, it is
necessary to analyze certain rare phenomena within such models in order to
test or at least determine the ballpark of model parameters. One such rare
decay is the radiative decay of $B$ mesons, $B\rightarrow X_s \gamma$ which
is under intense experimental investigation at the present $B$ factories.

The flavor mixing in the quark and lepton sectors will lead to flavor
changing (nondiagonal) couplings of the heavy $Z^{^{\prime }}$. Since its
topology is similar to the photon and gluon vertex, calculating the
branching ratio (BR) and CP asymmetry $(A_{CP})$ of the process $%
B\rightarrow X_{s}\gamma $ can restrict the couplings $\xi $ of the $%
Z^{^{\prime }}$. Although in SM CP asymmetry A$_{CP}(b\rightarrow s\gamma )$
is less than 1\%, in the extension of the SM, due to CP violating couplings
large CP asymmetries are possible \cite{kagepr}. In particular, large
asymmetries arise naturally in models with enhanced chromomagnetic dipole
operators. There is also flavor violating Z couplings if there is Z-$%
Z^{^{\prime }}$ mixing. This is another motivation to search for flavor
changing neutral current (FCNC) effects.

The prediction for the $B\rightarrow X_{s}\gamma $ branching ratio is
usually obtained by normalizing the result for the corresponding decay rate
to that for the semileptonic decay rate, thereby eliminating a strong
dependence on the b-quark mass \cite{kagep}.

The branching fraction is obtained from the CLEO experiment \cite{cleo99} as

\begin{equation}
BR(b\rightarrow s\gamma )=(3.15\pm 0.35\pm 0.32\pm 0.26)\times 10^{-4}
\end{equation}

where the photon energy is $2.1<E_{\gamma }<2.7$ $GeV.$ Therefore the
branching fraction is between

\begin{equation}
2.0\times 10^{-4}<BR<4.5\times 10^{-4}
\end{equation}

Direct CP violation can lead to a difference between the rates for $%
b\rightarrow s\gamma $ and $\bar{b}\rightarrow \overline{s}\gamma $ giving
rise to a non-zero value for the CP asymmetry\cite{cleo00}

\begin{equation}
A_{CP}=\frac{\Gamma (b\rightarrow s\gamma )-\Gamma (\bar{b}\rightarrow
\overline{s}\gamma )}{\Gamma (b\rightarrow s\gamma )+\Gamma (\bar{b}%
\rightarrow \overline{s}\gamma )}.
\end{equation}

The SM predicts that this asymmetry is very small, less than 1\%. Recent
theoretical work suggests that non-SM physics may contribute significantly
to a CP asymmetry, as large as 10-40\%.

The signature for $b\rightarrow s\gamma $ is a photon with energy
sufficiently high that it is unlikely to come from other B decay processes.

In Cleo's work the photon energy range is taken between 2.1-2.7 GeV, and in
conclusion CP asymmetry $A_{CP}$ in $b\rightarrow s\gamma $ plus $%
b\rightarrow d\gamma $ lies between the limits \cite{cleo00},

\begin{equation}
-0.27<A<+0.10
\end{equation}

at 90\% confidence level.

These limits rule out some extreme non-SM predictions, but are consistent
with most, as well as with the SM.

By using the constrained couplings of $Z^{^{\prime }}$ boson from this
process, we may calculate other processes including $Z^{^{\prime }}$ boson
couplings.

\section{Calculation}

At the W-boson mass scale Q=M$_{W},$ the flavor changing radiative
transition $b\rightarrow s\gamma $ is described by the following operator
product expansion\cite{buch,aok,grin}

\begin{eqnarray}
H_{eff} &=&-\frac{4G_{F}}{\sqrt{2}}K_{ts}^{\ast} K_{tb}\Big\{%
C_{2}(M_{W})O_{2}(M_{W})  \nonumber \\
&&+C_{7}(M_{W})O_{7}(M_{W})+C_{8}(M_{W})O_{8}(M_{W})\Big\}
\end{eqnarray}

where $C_{i}(M_{W})$ are the Wilson coefficients, and $O_{i}(M_{W})$ are the
local operators defined by

\begin{equation}
O_{2}(M_{W})=(\bar{s}_{L}\gamma _{\mu }c_{L})(\bar{c}_{L}\gamma _{\mu
}b_{L}),
\end{equation}

\begin{equation}
O_{7}(M_{W})=\frac{e(Q)}{16\pi ^{2}}\bar{m}_{b}(M_{W})(\bar{s}_{L}\sigma
^{\mu \nu }b_{R})F_{\mu \nu },
\end{equation}

\begin{equation}
O_{8}(M_{W})=\frac{g_{s}(Q)}{16\pi ^{2}}\bar{m}_{b}(M_{W})(\bar{s}_{L}\sigma
^{\mu \nu }T^{a}b_{R})G_{\mu \nu }^{a}.
\end{equation}

Here $O_{2}(M_{W})$ is a four-fermion operator and its coefficient satisfies
$C_{2}(M_{W})=1$ $,$ that is, it is a constant independent of the looping
particle species. However, the coefficients of the electric $O_{7}(M_{W})$
and color $O_{8}(M_{W})$ dipole operators can receive nonvanishing
contributions from physics beyond the SM. In general

\begin{equation}
C_{7,8}(M_{W})=C_{7,8}^{SM}(M_{W})+C_{7,8}^{NP}(M_{W})
\end{equation}

where the last term is for the New Physics contributions, and the SM
contributions $C_{7,8}^{SM}(M_{W})$ are \cite{chet};

\begin{equation}
C_{7}^{SM}(M_{W})=\frac{3x^{3}-2x^{2}}{4(x-1)^{4}}\ln x+\frac{%
-8x^{3}-5x^{2}+7x}{24(x-1)^{3}}
\end{equation}

\begin{equation}
C_{8}^{SM}(M_{W})=\frac{-3x^{2}}{4(x-1)^{4})}\ln x+\frac{-x^{3}+5x^{2}+2x}{%
8(x-1)^{3}}.
\end{equation}

Here

\begin{equation}
x=\frac{m_{t,p}^{2}}{M_{W}^{2}}\left( \frac{\alpha _{s}(M_{W})}{\alpha
_{s}(m_{b})}\right) ^{24/23}\left( 1-\frac{8}{3}\frac{\alpha _{s}(m_{t})}{%
\pi }\right) .
\end{equation}

In the Langacker-Pl\"{u}macher paper \cite{plum}, we let $\xi _{R}^{sb}=\xi
_{L}^{sb}\equiv \xi _{sb}$ for simplicity. Then the new physics contribution
can be parametrized as

\begin{equation}
C_{7,8}^{NP}(M_{W})=2\sqrt{2}\frac{\xi _{sb}}{V_{ts}^{\ast }V_{tb}}
\end{equation}

where we have used the fact that gluon and photon diagrams have the same
topology as $Z^{^{\prime }}$ boson which is neutral both electrically and
chromoelectrically. Here $\xi _{sb}$ is a complex number whose expression is
\cite{plum}

\begin{equation}
\xi _{sb}=\frac{y}{m_{b}}(B^{d}m_{d}B^{d})_{23}+w\epsilon _{L}(b)B_{23}^{d}
\end{equation}

where $m_{d}$ is the diagonal mass matrix of down quarks,

\begin{equation}
y=\left( \frac{g_{2}}{g_{1}}\right) ^{2}(\rho _{1}\sin ^{2}\theta +\rho
_{2}\cos ^{2}\theta ),
\end{equation}

\begin{equation}
w=\frac{g_{2}}{g_{1}}\sin \theta \cos \theta (\rho _{1}-\rho _{2}),
\end{equation}

\begin{equation}
\rho _{i}=\frac{M_{W}^{2}}{M_{i}^{2}\cos ^{2}\theta _{W}}.
\end{equation}

$M_{i}$ are the masses of the neutral gauge boson mass eigenstates, $g_{1}$
and $g_{2}$ are the coupling constants of the neutral gauge bosons Z and Z$%
^{^{\prime }}$ respectively.

The Wilson coefficients mentioned so far are at the $Q=M_{W}$ scale.
However, physically one must recalculate them at the $Q=m_{b}$ scale- the
natural scale of the problem. Therefore, the Wilson coefficients above are
now reduced to $Q=m_{b}$ scale via\cite{kagep}

\begin{eqnarray}
C_{2}(m_{b}) &=&\frac{1}{2}(\eta ^{-\frac{12}{23}}+\eta ^{\frac{6}{23}}),
\nonumber \\
C_{7}(m_{b}) &=&\eta ^{\frac{16}{23}}C_{7}(M_{W})+\frac{8}{3}(\eta ^{\frac{14%
}{23}}-\eta ^{\frac{16}{23}})C_{8}(M_{W})+C_{2}(M_{W})\sum%
\limits_{i=1}^{8}h_{i}^{(72)}\eta ^{a_{i}},  \nonumber \\
C_{8}(m_{b}) &=&\eta ^{\frac{14}{23}}C_{8}(M_{W})+C_{2}(M_{W})\sum%
\limits_{i=1}^{8}h_{i}^{(82)}\eta ^{a_{i}},
\end{eqnarray}

where the magic numbers on the right are given by

\begin{eqnarray}
a_{i} &=&(\frac{14}{23},\frac{16}{23},\frac{6}{23},-\frac{12}{23}%
,0.4086,-0.4230,-0.8994,0.1456),  \nonumber \\
h_{i}^{(72)} &=&(\frac{626126}{272277},-\frac{56281}{51730},-\frac{3}{7},-%
\frac{1}{14},  \nonumber \\
&&-0.6494,-0.0380,-0.0186,-0.005),  \nonumber \\
h_{i}^{(82)} &=&(\frac{313063}{363036},0,0,0,-0.9135,0.0873,-0.0571,0.0209).
\end{eqnarray}

In obtaining $C_{7}(m_{b})$ we have made use of the leading order QCD
renormalization group running from $Q=M_{W}$ down to $Q=m_{b}$ and so the
renormalization factor $\eta =\alpha _{s}(M_{W})/\alpha _{s}(m_{b}).$ Having
$C_{7}(m_{b})$ at hand, it is easy to calculate the branching ratio and CP
asymmetry of the decay. Branching ratio is given by

\begin{equation}
BR(b\rightarrow s\gamma )=BR^{(\exp )}(b\rightarrow ce\overline{\nu }%
_{e})\left| \frac{V_{ts}^{\ast }V_{tb}}{V_{cb}}\right| ^{2}\frac{6\alpha }{%
\pi f(z_{c})}S(\delta )\left| C_{7}(m_{b})\right| ^{2}  \label{br}
\end{equation}

where $z_{c}=m_{c}^{2}/m_{b}^{2},$ $BR^{(\exp )}(B\rightarrow X_{c}e%
\overline{\nu }_{e})\approx 10.5\%,$ and $f(z)=1-8z+8z^{3}-z^{4}-12z^{2}\ln
z.$ Here $\delta =0.9$ represents the energy of the emitted photon, and

\begin{equation}
S(\delta )=\exp \left[ -\frac{2\alpha _{s}(m_{b})}{3\pi }\left( \ln
^{2}\delta +\frac{7}{2}\ln \delta \right) \right] .
\end{equation}

Similarly the CP asymmetry is given by

\begin{eqnarray}
A_{CP}(b& \rightarrow s\gamma )=\frac{\alpha _{s}(m_{b})}{\left|
C_{7}(m_{b})\right| ^{2}}\{\frac{40}{81}\mbox{Im}\left[ C_{2}(m_{b})C_{7}^{%
\ast }(m_{b})\right]  \nonumber \\
& -\frac{8z_{c}}{9}\left[ v(z_{c})+b(z_{c},\delta )\right] \mbox{Im}\left[
\left( 1+\epsilon _{s}\right) C_{2}(m_{b})C_{7}^{\ast }(m_{b})\right]
\nonumber \\
& -\frac{4}{9}\mbox{Im}\left[ C_{8}(m_{b})C_{7}^{\ast }(m_{b})\right]
\nonumber \\
& +\frac{8z_{c}}{27}b(z_{c},\delta )\mbox{Im}\left[ \left( 1+\epsilon
_{s}\right) C_{2}(m_{b})C_{8}^{\ast }(m_{b})\right] \}  \label{acp}
\end{eqnarray}

where $\epsilon _{s}$ represents the pure SM contribution

\begin{equation}
\epsilon _{s}\equiv \frac{V_{us}^{\ast }V_{ub}}{V_{ts}^{\ast }V_{tb}}\approx
\lambda _{c}^{2}(i\eta -\rho )=O(10^{-2}),
\end{equation}

and the functions $v(z)$ and $b(z,\delta )$ are defined by

\begin{eqnarray}
g(z,y) &=&\theta (y-4z)\{(y^{2}-4yz+6z^{2})\ln \left( \sqrt{\frac{y}{4z}}+%
\sqrt{\frac{y}{4z}-1}\right)  \nonumber \\
&&-\frac{3y(y-2z)}{4}\sqrt{1-\frac{4z}{y}}\},
\end{eqnarray}

\begin{eqnarray}
v(z) &=&\left( 5+\ln z+\ln^{2}{z}-\frac{\pi^{3}}{3}\right) +(\ln^{2}{z}-%
\frac{\pi^{3}}{3})z+  \nonumber \\
&&\left(\frac{28}{9}-\frac{4}{3}\ln z\right) z^{2}+O(z^{3}),
\end{eqnarray}
with $b(z,\delta )=g(z,1)-g(z,1-\delta ).$

\section{Results}

To get the numerical results from Eqs. \ref{br}. and \ref{acp}. we use $%
\delta =0.90.$ We calculate the branching ratio from Eq. (\ref{br}) as
\begin{equation}
BR(b\rightarrow s\gamma )=0.00025+0.01232\ast I^{2}-0.00352\ast
R+0.01232\ast R^{2}
\end{equation}

The branching ratio $BR(b\rightarrow s\gamma )$ is plotted in Fig. 1 as a
function of $R$ for different values of $I.$

Comparing the graph in Fig 1. with the experimental value \cite{cleo99}, we
get the following values with $\xi =R+i\ast I$

\begin{eqnarray}
-0.03 &<&R<+0.02\mbox{ for }I=0.04,  \nonumber \\
-0.02 &<&R<+0.04\mbox{ for }I=0.08,  \nonumber \\
+0.01 &<&R<+0.11\mbox{ for }I=0.12
\end{eqnarray}

to get the following constraint

\begin{equation}
0.000222<BR(b\rightarrow s\gamma )<0.000408.
\end{equation}

From the values obtained for R and I the coupling constant $\xi =R+i\ast I$
is between

\begin{equation}
0.05\leq |\xi _{sb}|\leq 0.163.
\end{equation}

We calculate the CP asymmetry $A_{CP}(b\rightarrow s\gamma )$ from Eq. (\ref
{acp}) as

\begin{equation}
A_{CP}(b\rightarrow s\gamma )=\frac{-0.00648Re[I]}{0.0963452+4.71401\,{{I}%
^{2}}-1.34785\,R+4.71401\,{R^{2}}}
\end{equation}

\medskip

The CP asymmetry $A_{CP}(b\rightarrow s\gamma )$ is plotted in Fig. 2 as a
function of $R$ for different values of $I.$

Similarly From Fig. 2, we obtain

\begin{equation}
-0.13<R<+0.15\mbox{ for }I=\pm 0.002
\end{equation}

to get the following experimental constraint

\begin{equation}
-0.27<A_{CP}(b\rightarrow s\gamma )<+0.10,
\end{equation}

from \cite{cleo00}.

From the values obtained for R and I the coupling constant $\xi =R+i\ast I$
is between

\begin{equation}
0.132\leq \xi _{sb}\leq 0.151.
\end{equation}

\section{Discussion and Conclusion}

In summary we have analyzed the constraints on family non-universal $%
Z^{\prime }$ couplings from the branching ratio and the CP asymmetry of the
decay $B\rightarrow X_{s}\gamma $. We have found that the parameter $\xi
_{sb}$ has non-vanishing real and imaginary parts, and the central value
allowed is $|\xi _{sb}|\sim 0.1$. Although it is not possible to infer any
further results about the parameters $\xi _{sb}$ depends on, still one can
infer that the parameter $\xi _{sb}$ is complex and is required to be around
$0.1$ in magnitude.

As shown in Fig. 1 the dependence of the branching ratio on $R$ is strong,
and given the present $1\sigma$ bounds we conclude that only positive values
of $R$ are preferred. Similarly, the graph in Fig. 2 depicts the CP
asymmetry of the decay, and it can be as large as $10\%$ in the parameter
region preferred by the branching ratio constraints.

Though we have restricted our analysis to $B\rightarrow X_s \gamma$ only, it
is clear that such family--non-universal $Z^{\prime}$ bosons will contribute
to various FCNC observables. Among others, the two hadronic decays $%
B\rightarrow J\psi K_s$ and $B\rightarrow \pi K_s$ are of prime importance.
The CP violation in the former has already been measured constituting the
present value of $\sin2\beta$. The latter, however, is a pure penguin
process and it is still under investigation. Any measurable difference
between the CP asymmetries of respective decay modes will be a violation of
the SM expectation. This then can be taken as a signal of the new physics
effects, among all possible candidates, the family non--universal $%
Z^{\prime} $ models are of particular importance since any difference
between the couplings to charm and strange quarks will show up as shift from
$\sin 2\beta $.

\newpage

FIGURE CAPTIONS

\noindent

Figure 1. Plot of branching ratio (BR) as a function of R for different
values of I.

\noindent

Figure 2. Plot of CP asymmetry $A_{CP}$ as a function of R for different
values of I.

\begin{figure}[tbph]
\unitlength1mm
\begin{picture}(161,50)
\put(5,-5){\psfig{file=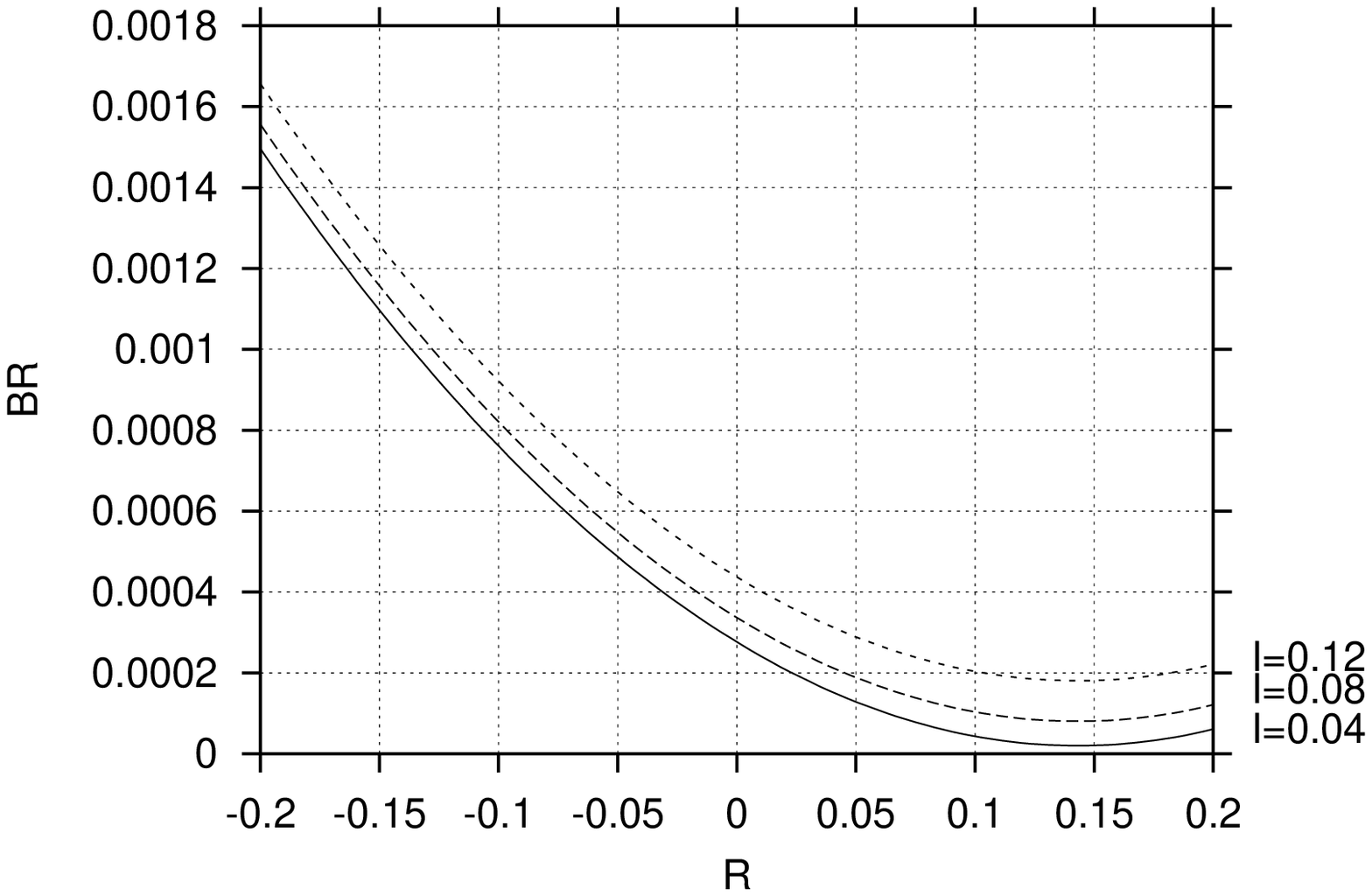,width=10cm,angle=270} }
\end{picture}
\end{figure}

\newpage

\begin{figure}[tbph]
\unitlength1mm
\begin{picture}(161,50)
\put(5,-5){\psfig{file=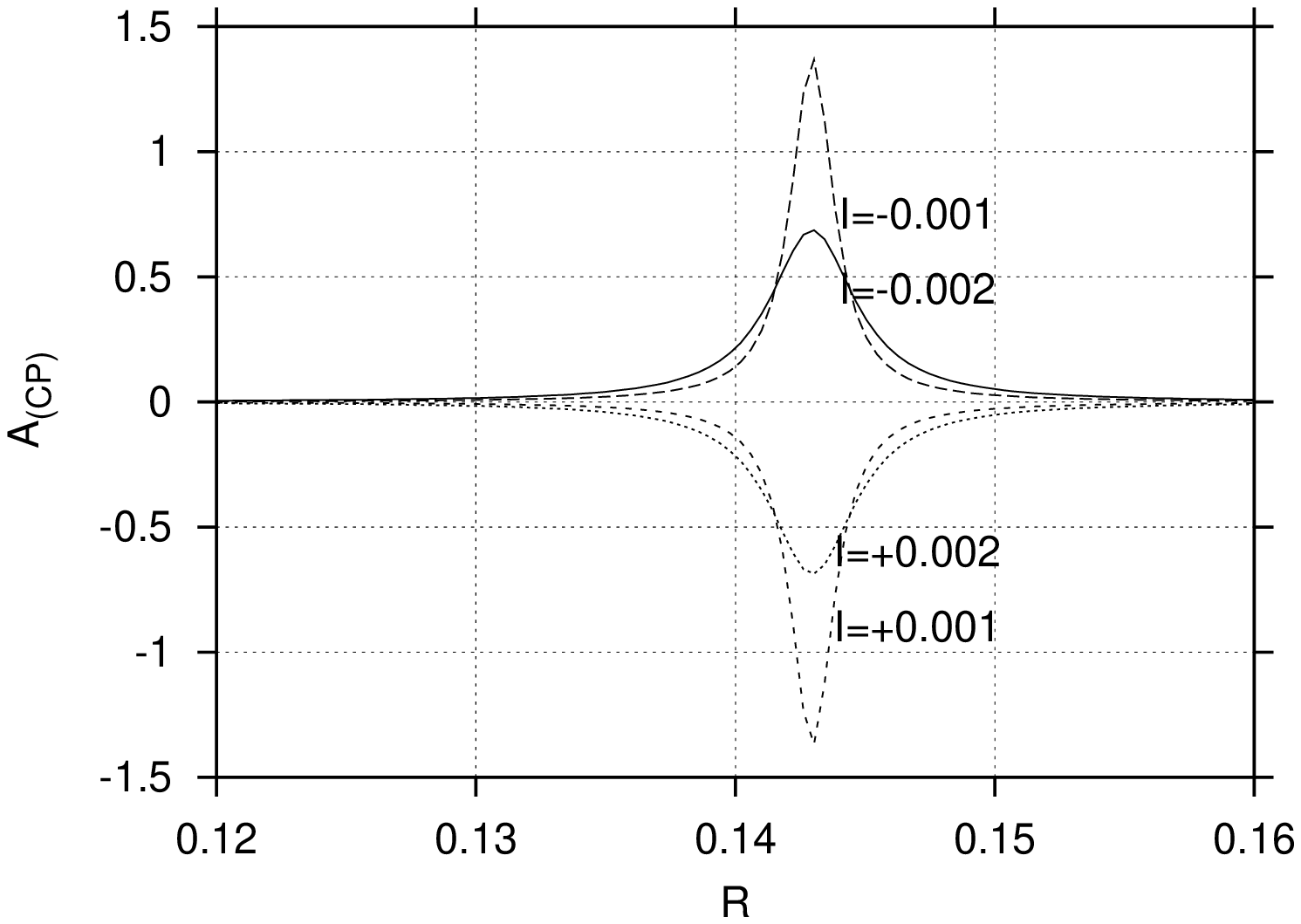,width=10cm,angle=270} }
\end{picture}
\end{figure}

\newpage

\end{document}